\documentclass[preprintnumbers,twocolumn,article,amsmath,amssymb,floatfix,9pt,prd,superscriptaddress,nofootinbib]{revtex4-1}
\usepackage[utf8,latin1]{inputenc}
\usepackage{graphicx}
\usepackage{dcolumn}
\usepackage[dvipsnames]{xcolor}
\usepackage[T1]{fontenc}

\usepackage{mathrsfs}  
\usepackage{cases}
\usepackage{bm}
\usepackage{academicons}
\usepackage{mathtools, nccmath}
\usepackage{fancyhdr}
\usepackage{tikz,xcolor}

\usepackage{tensor}
\usepackage[normalem]{ulem}
\usepackage{lipsum}
\usepackage{soul}
\usepackage{cancel}
\usepackage{stackengine,scalerel}
\usepackage{mathabx}
\usepackage{hyperref}
\usepackage{tabularx}
\hypersetup{colorlinks, linkcolor={red},citecolor={blue},urlcolor={blue}}  

\AtBeginDocument{\fontsize{8}{11}\selectfont} 

\definecolor{lime}{HTML}{A6CE39}
\DeclareRobustCommand{\orcidicon}{
	\begin{tikzpicture}
	\draw[lime, fill=lime] (0,0) 
	circle [radius=0.16] 
	node[white] {{\fontfamily{qag}\selectfont \tiny ID}};
	\draw[white, fill=white] (-0.0625,0.095) 
	circle [radius=0.007];
	\end{tikzpicture}
	\hspace{-2mm}
}

\fancypagestyle{plain}{%
  \fancyhf{}
  \fancyfoot[C]{\iffloatpage{}{\thepage}}
  }
\foreach \x in {A, ..., Z}{%
	\expandafter\xdef\csname orcid\x\endcsname{\noexpand\href{https://orcid.org/\csname orcidauthor\x\endcsname}{\noexpand\orcidicon}}
}

\begin{document}

\title[EHT-based observational constraints on regular black holes in higher dimensions]{Event Horizon Telescope Observational Constraints on Dymnikova-Type Non-Singular Black Holes in Higher Dimensions}

\author{A. Errehymy\orcidA{}}%
\email{abdelghani.errehymy@gmail.com}
\affiliation{Astrophysics Research Centre, School of Mathematics, Statistics and Computer Science, University of KwaZulu-Natal, Private Bag X54001, Durban 4000, South Africa}
\affiliation{Center for Theoretical Physics, Khazar University, 41 Mehseti Str., Baku, AZ1096, Azerbaijan}

\author{Y. Khedif\orcidB{}}%
\email{youssef.khedif@gmail.com}
\affiliation{Laboratory of Mechanics and High Energy Physics, Department of Physics, Faculty of Sciences A\"in Chock, University Hassan II, P.O. Box 5366, Maarif Casablanca, 20100, Morocco}

\author{M. Daoud\orcidD{}}%
\email{m$_{}$daoud@hotmail.com}
\affiliation{Department of Physics, Faculty of Sciences, Ibn Tofail University, P.O. Box 133, Kenitra 14000, Morocco}
\affiliation{Abdus Salam International Centre for Theoretical Physics, Miramare, Trieste 34151, Italy}

\author{Y. Myrzakulov\orcidC{}}%
\email{ymyrzakulov@gmail.com} 
\affiliation{Department of General and Theoretical Physics, L.N. Gumilyov Eurasian National University, Astana 010008, Kazakhstan}

\author{O. Donmez\orcidE{}}%
\email{orhan.donmez@aum.edu.kw}
\affiliation{College of Engineering and Technology, American University of the Middle East, Egaila 54200, Kuwait}

\author{B. Turimov\orcidD{}}%
\email[]{bturimov@astrin.uz}
\affiliation{Central Asian University, Milliy bog Str. 264, Tashkent, 111221, Uzbekistan}
\affiliation{University of Tashkent for Applied Sciences, Gavhar Str. 1, Tashkent, 100149, Uzbekiston}

\date{\today} 

\begin{abstract}
{\footnotesize {Black holes are among the most compelling predictions of general relativity (GR) and are now strongly supported by observations from gravitational-wave detectors and the Event Horizon Telescope (EHT). While standard black hole solutions suffer from central singularities, regular black holes avoid this issue by introducing a nonsingular core. In this work, we extend the Dymnikova regular black hole to higher dimensions using a smooth matter distribution. The resulting spacetime features a de Sitter-like core and two horizons. We analyze photon motion and show that circular photon orbits remain unstable, giving rise to a well-defined black hole shadow. Our results indicate that the shadow size grows with the black hole scale but decreases slightly as the number of dimensions increases. We also investigate thermodynamic properties, including Hawking temperature and energy emission, and find a strong dependence on dimensionality. Finally, we compare our model with EHT observations to place constraints on the parameters and highlight potential observational signatures of higher-dimensional regular black holes.}}
\end{abstract}

\maketitle
\textbf{Introduction:} Among the most captivating predictions of Einstein's relativity are black holes. Their presence is supported by a variety of observations: gravitational waves from colliding compact objects detected by the LIGO-Virgo-KAGRA observatories~\cite{LIGOScientific:2025kei, LIGOScientific:2016aoc}; the remarkable images of M87$^{*}$ and Sgr A$^{*}$ black hole shadows captured by the EHT~\cite{EventHorizonTelescope:2022wkp, EventHorizonTelescope:2019dse}; and the radiation emitted by matter spiraling in their accretion disks~\cite{Nampalliwar:2020tup}. Black holes reveal a central feature of GR: the existence of singularities. Schwarzschild's 1916 solution to Einstein's field equations, describing the spacetime around a static, spherically symmetric mass, laid the groundwork for black hole theory. This solution shows two prominent characteristics: a central singularity, where curvature invariants like the Kretschmann scalar diverge as $r \to 0$, and an event horizon located at $r = 2M$, marking a one---way boundary from which nothing, including light, can escape.

For a long time, singularities in spacetime---like those appearing in Schwarzschild, Oppenheimer-Snyder, or Kerr solutions \cite{poisson2004relativist}---were viewed as mathematical curiosities resulting from idealized symmetries, and many expected them to vanish in realistic conditions. This view was overturned by the singularity theorems formulated by Penrose and Hawking \cite{hawking2023large}. In particular, Penrose's 1965 theorem \cite{Penrose:1964wq} showed that singularities are an unavoidable feature of gravitational collapse, emerging inevitably once a closed trapped surface forms \cite{Landsman:2022hrn, Landsman:2021mjt}. Such surfaces are so intensely warped that even outgoing light cannot escape, being forced to converge \cite{hawking2023large}. While singularity theorems do not directly prove that black holes exist, they reveal a key feature of spacetime: under reasonable conditions, causal geodesics cannot be extended indefinitely along their affine parameters. To interpret this incompleteness as the formation of a black hole, the cosmic censorship conjecture is invoked \cite{Penrose:1969pc}. Proposed by Penrose, this conjecture suggests that any singularity produced by gravitational collapse is always hidden within a black hole, making it invisible to distant observers \cite{wald1984general}.

The presence of singularities points to a fundamental breakdown in GR, where the theory cannot provide meaningful boundary conditions for the field equations \cite{hawking1979path}. In these regions, classical physics fails: physical quantities become infinite, observations are impossible, and the theory loses predictive power. This indicates that, while highly successful in many domains, GR may not be the ultimate theory of gravity. Coupled with recent observations suggesting alternatives to standard black holes \cite{Herdeiro:2022yle, LISA:2022kgy, Barack:2018yly}, this has motivated exploration into modified and extended gravity models that address these shortcomings while retaining Einstein,s tested predictions \cite{Capozziello:2011et}. {More broadly, recent work on Lorentzian-Euclidean black holes and quantum-modified spacetimes \cite{Capozziello:2024ucm, Capozziello:2025wwl, DeBianchi:2025bgn, Wang:2025fmz} suggests that shifts in the causal structure, the appearance of atemporal regions, and quantum effects can naturally remove singularities and lead to noticeable changes in observable quantities, such as black hole shadows and the behavior of accreting matter.} In addition to horizonless exotic compact objects capable of reproducing black-hole-like behavior \cite{Cardoso:2019rvt}, several regular black hole models have been proposed. These include higher-dimensional solutions from string theory, such as Myers-Perry black holes, which avoid singularities in specific odd dimensions and spin setups for $r \ge 0$ \cite{Myers:2011yc, Emparan:2008eg}, as well as numerous four-dimensional constructs reviewed in \cite{Carballo-Rubio:2025fnc, Lan:2023cvz}.

The first example of a static, spherically symmetric regular black hole with an event horizon was introduced by Bardeen \cite{bardeen1968non}, who replaced the constant Schwarzschild mass with an $r$-dependent mass function. Building on early ideas by Sakharov \cite{sakharov1966initial} and Gliner \cite{gliner1966algebraic}, Dymnikova later proposed a nonsingular Schwarzschild-like geometry, now known as the Dymnikova black hole \cite{Dymnikova:1992ux}. This model, sourced by a spherically symmetric vacuum, possesses a core that becomes asymptotically de Sitter as $r \to 0$. Inspired by these works, additional spherically symmetric regular black hole solutions have been developed \cite{dymnikova2003spherically, Hayward:2005gi}, with their thermodynamic properties discussed extensively in \cite{Macedo:2024dqb, Fathi:2021liw, Molina:2021hgx}. On the other hand, the gravitational analogue of the Schwinger effect, first introduced by Dymnikova \cite{Dymnikova:1992ux}, describes a vacuum solution that smoothly connects two regimes: at large radii it behaves like the Schwarzschild geometry, while near the center it approaches a de Sitter core. Over the past years, this so-called Dymnikova black hole has been explored in different theoretical frameworks, including pure Lovelock gravity \cite{Estrada:2024uuu}, the generalized uncertainty principle \cite{Ma:2024tqp}, higher-curvature corrections \cite{Konoplya:2024kih}, and even wormhole constructions \cite{Estrada:2023pny}. Its quasinormal spectra and thermodynamic behavior in higher dimensions have also been extensively investigated \cite{Macedo:2024dqb}.

The observation of gravitational waves by the LIGO and Virgo collaborations \cite{LIGOScientific:2017ycc, LIGOScientific:2016sjg, LIGOScientific:2016kms} has marked a turning point in cosmology, opening possibilities that were previously out of reach. These signals now provide an invaluable means of probing the universe, including applications to gravitational lensing studied in the weak-field limit \cite{Mukherjee:2019wcg, Contigiani:2020yyc}. Early investigations into lensing emphasized the bending of light across cosmic distances, usually modeled with the Schwarzschild geometry \cite{darwin1959gravity}, and were later generalized to encompass static, spherically symmetric spacetimes \cite{atkinson1965light}. In regions of intense gravity, however---most notably near black holes---the deviation of light is far stronger, in full agreement with expectations from strong-field gravity. The historic achievement of the EHT in capturing the image of the supermassive black hole at the core of M87 drew immense attention across the scientific community \cite{EventHorizonTelescope:2019ggy, EventHorizonTelescope:2019pgp, EventHorizonTelescope:2019uob, EventHorizonTelescope:2019ths}. Long before this observational milestone, Virbhadra and Ellis \cite{Virbhadra:1999nm} had devised a simplified lens equation to investigate how supermassive black holes in asymptotically flat spacetimes influence the path of light. Their work demonstrated that the overwhelming gravitational pull near such objects can give rise to multiple images, arranged symmetrically along the optical axis---a striking signature of strong-field lensing. Recent theoretical and computational advances naturally complement these developments. Hamiltonian and dynamical methods help clarify geodesic motion and stability in black hole spacetimes \cite{qiao2025calculation}, while pulsar timing and time-delay techniques are closely tied to relativistic signal propagation in strong gravitational fields \cite{xie2025time}. Reliable, energy-stable numerical schemes remain essential for solving the nonlinear equations governing such systems \cite{bo2022discrete}, and progress in mission analysis and propulsion concepts may open new possibilities for future space-based tests of gravity near compact objects \cite{huo2023fast}.

Building on this foundation, further progress was made through the analytical contributions of Fritelli et al. \cite{Frittelli:1999yf}, Bozza et al. \cite{Bozza:2001xd}, and Tsukamoto \cite{Tsukamoto:2016jzh}, who refined techniques for studying gravitational lensing in the strong-field regime. Their work broadened the scope of investigation, addressing light deflection not only in Schwarzschild spacetime \cite{Virbhadra:1998dy, Virbhadra:2002ju, Oguri:2019fix, Ezquiaga:2020gdt, Okyay:2021nnh, Li:2020dln} but also in more exotic scenarios such as wormholes \cite{Tsukamoto:2012xs, Tsukamoto:2016qro, Tsukamoto:2017edq, Ovgun:2018xys}, rotating black holes \cite{Hsieh:2021rru, Hsieh:2021scb}, and models inspired by alternative theories of gravity \cite{Chakraborty:2016lxo, AraujoFilho:2024mvz}. Lensing phenomena in Reissner-Nordstr\"om geometries \cite{Tsukamoto:2016oca, Eiroa:2003jf, Eiroa:2002mk} and other related configurations \cite{Tsukamoto:2022uoz, Zhang:2024sgs} have likewise been examined, further enriching our understanding of how extreme gravitational fields shape the propagation of light. 

Motivated by these arguments, in this work, we shall investigate the higher-dimensional extension of the Dymnikova black hole and examine its physical as well as observational properties. Starting from a density profile inspired by the gravitational analogue of the Schwinger mechanism, we shall construct the associated spacetime metric and follow its generalization to higher dimensions. Our analysis shall address the horizon structure, the definition of mass, and the thermodynamical behavior of the solution, with particular emphasis on the Hawking temperature and the corresponding energy emission rate. To test the regularity of the geometry, we shall evaluate curvature invariants, while the motion of photons shall be employed to study photon spheres and the propagation of light in the surrounding spacetime. Building on these results, we shall characterize the shadow of the black hole, describing its angular size and celestial coordinates, and highlight the signatures imprinted by higher-dimensional effects and the role of the Dymnikova parameters. Taken together, these investigations shall provide a comprehensive framework that links the theoretical construction of regular black holes with their potential astrophysical manifestations in extended spacetime scenarios.\\

%-----------------------------------------------------------------------
\textbf{The Dymnikova black hole metric in higher-dimensional spacetime:}
%-----------------------------------------------------------------------
The Dymnikova black hole provides a regularized alternative to the classical Schwarzschild solution. Originally formulated in four dimensions, its density profile can be interpreted as a gravitational analogue of the Schwinger mechanism in quantum electrodynamics \cite{Dymnikova:1996gob}, where strong electric fields produce particle-antiparticle pairs. In the gravitational setting, the energy density of the vacuum is spread over a finite region, effectively smoothing the central singularity.\\

%---------------------------------------------------------
\textbf{Coordinate transformation and density profile:}
%---------------------------------------------------------
Following \cite{Macedo:2024dqb, Paul:2023pqn}, a characteristic length scale $r_{\star}$ is introduced via
%\begin{equation}
$\frac{E_0}{E} = \frac{r^3}{r_{\star}^3}, \quad r_{\star}^3 = r_s r_0^2$,
%\end{equation}
with $E \sim r^{-3}$, $E_0 = \frac{\pi \hbar m_e^2}{e}$, $r_s$ the Schwarzschild radius, and $r_0$ the de Sitter core scale. This leads to a smooth radial density profile
\begin{equation}
\rho(r) = \rho_0 \exp\Big(-\frac{r^3}{r_{\star}^3}\Big), \qquad r_0^2 = \frac{3}{8\pi \rho_0}.
\end{equation}

%-------------------------------------------
\textbf{Extension to higher dimensions:}
%-------------------------------------------
In $D$ dimensions, the characteristic scale is generalized as
\begin{equation}
r_{\star}^{D-1} = r_0^2 r_s^{D-3}, \qquad \mathcal{F}(r) \equiv e^{-r^{D-1}/r_{\star}^{D-1}}.
\end{equation}
The corresponding stress-energy tensor retains spherical symmetry:
%\begin{equation}
$T^\mu{}_\nu = \text{diag}(-\rho, P_r, P_t, \ldots)$,
%\end{equation}
with
%\begin{equation}
$T^0{}_0 = T^1{}_1, \quad T^{\theta_i}{}_{\theta_i} = \dots = T^{\theta_{D-2}}{}_{\theta_{D-2}}$.
%\end{equation}
The $D$-dimensional energy density then becomes
\begin{equation}
\rho(r) = -T^0{}_0 = \rho_0 \mathcal{F}(r).
\end{equation} 

%---------------------------------------------
\textbf{Metric potential and line element:}
%---------------------------------------------
The resulting lapse function is
\begin{equation}\label{laps}
f(r) = 1 - \frac{r_s^{D-3}}{r^{D-3}} \big(1 - \mathcal{F}(r)\big),
\end{equation}
leading to the spacetime metric
\begin{equation}
ds^2 = -f(r) dt^2 + \frac{dr^2}{f(r)} + r^2 d\Omega_{D-2}^2,
\end{equation}
with $r_0^2 = \frac{(D-1)(D-2)}{16\pi \rho_0}$. The stress-energy components simplify to
\begin{eqnarray}
T^0{}_0 = T^1{}_1 &=& -\rho_0 \mathcal{F}(r), \\
T^{\theta_2}{}_{\theta_2} &=& \Big[\frac{D-1}{D-2} \Big(\frac{r}{r_{\star}}\Big)^{D-1} - 1\Big] \rho_0 \mathcal{F}(r).
\end{eqnarray}

%-----------------------------------------------------------------------------
\begin{figure*}
\centering
\includegraphics[width=5.9cm,height=4.9cm]{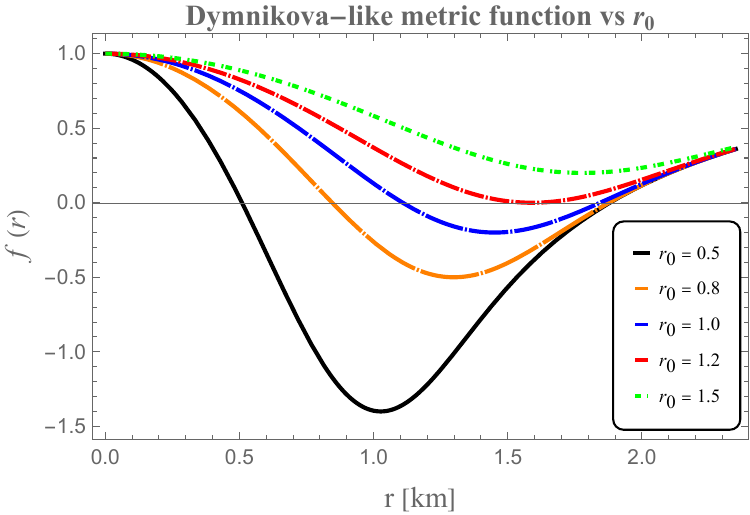}%
\includegraphics[width=5.9cm,height=4.9cm]{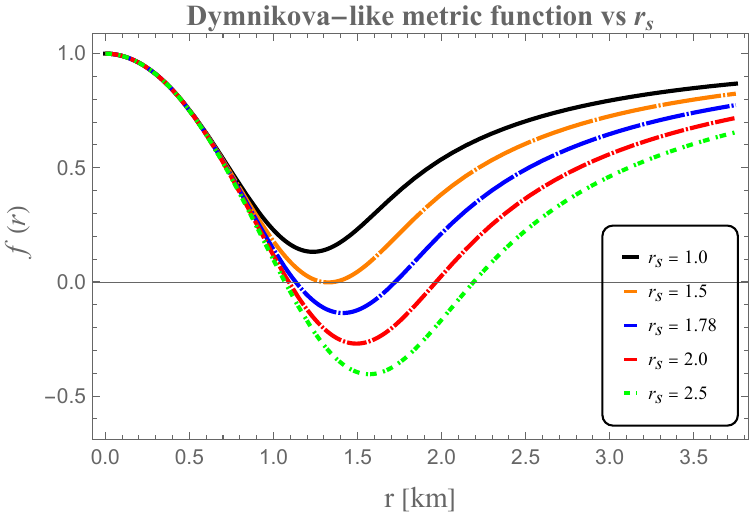}%
\includegraphics[width=5.9cm,height=4.9cm]{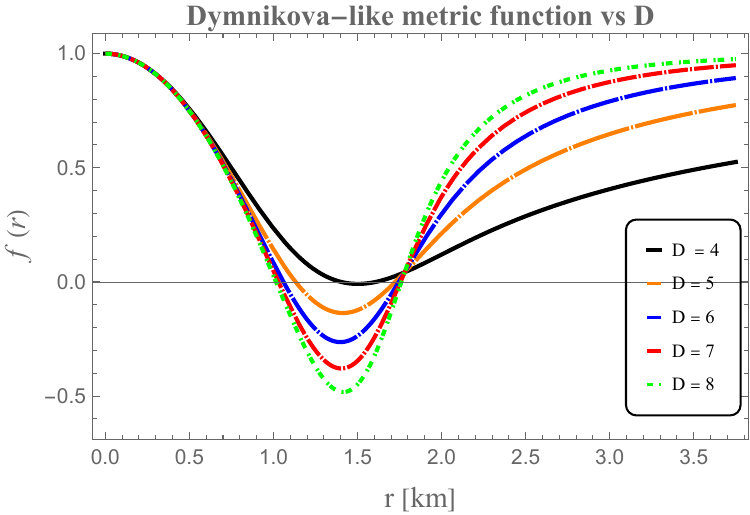}%
    \caption{\footnotesize The metric function $f(r)$ is examined across different parameter choices in higher-dimensional Dymnikova spacetimes. The left plot highlights how changing $r_0$ alters the profile while keeping $D=5$ and $r_s=1.88$ constant. The middle plot focuses on the influence of the dimensionality $D$, with $r_0=1.0$ and $r_s=1.88$. The right plot illustrates the impact of varying $r_s$ for fixed $r_0=1.0$ and $D=5$. }    \label{Fig1}
\end{figure*}
%-----------------------------------------------------------------------------
%-----------------------------------------------------------------------------
\begin{figure*}
\begin{center}
\includegraphics[width=5.9cm,height=4.9cm]{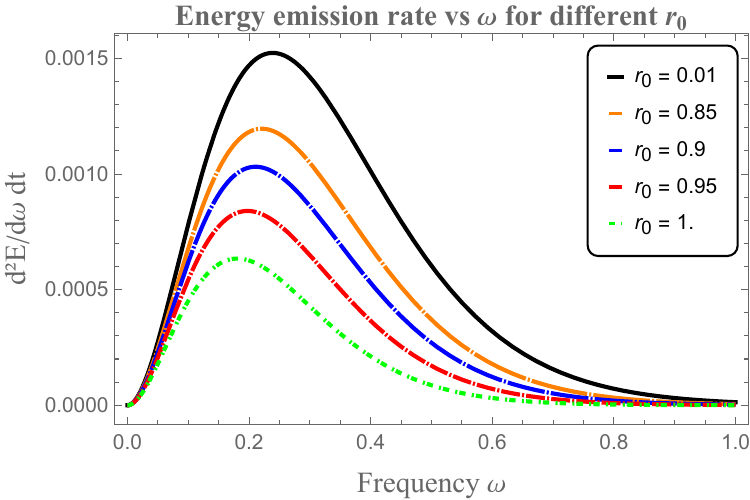}
\includegraphics[width=5.9cm,height=4.9cm]{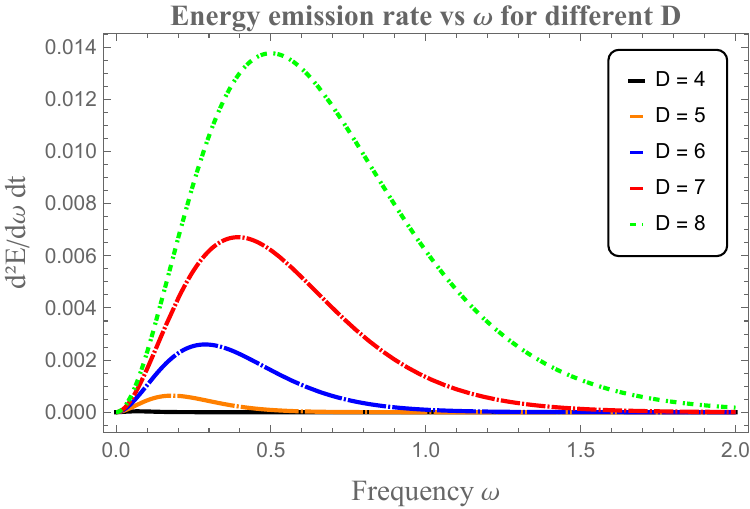}
\includegraphics[width=5.9cm,height=4.9cm]{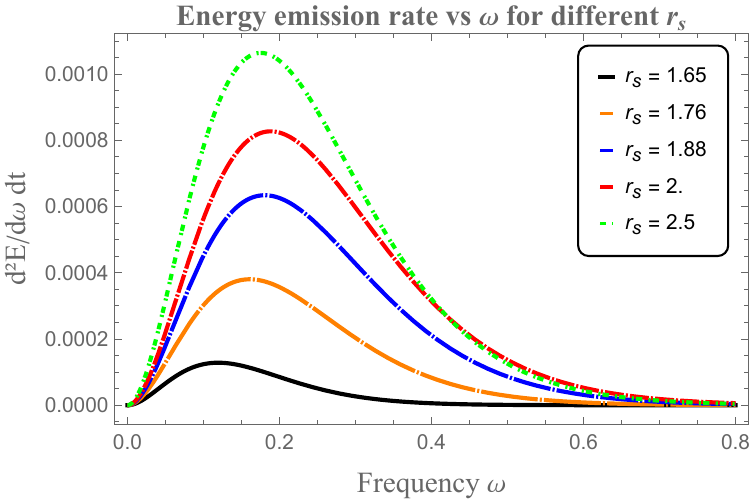}
\end{center}
\caption{\footnotesize The energy emission rate is investigated under different parameter choices in higher-dimensional Dymnikova spacetimes. The left panel shows the effect of varying $r_0$ while keeping $D=5$ and $r_s=1.88$ fixed. The middle panel illustrates the role of the dimensionality $D$, with $r_0=1.0$ and $r_s=1.88$. The right panel depicts the influence of changing $r_s$ for fixed $r_0=1.0$ and $D=5$. }\label{Fig6}
\end{figure*}
%---------------------------------------------------------------------------
%-----------------------------------------------------------------------------
\begin{figure*}
\begin{center}
\includegraphics[width=5.9cm,height=4.9cm]{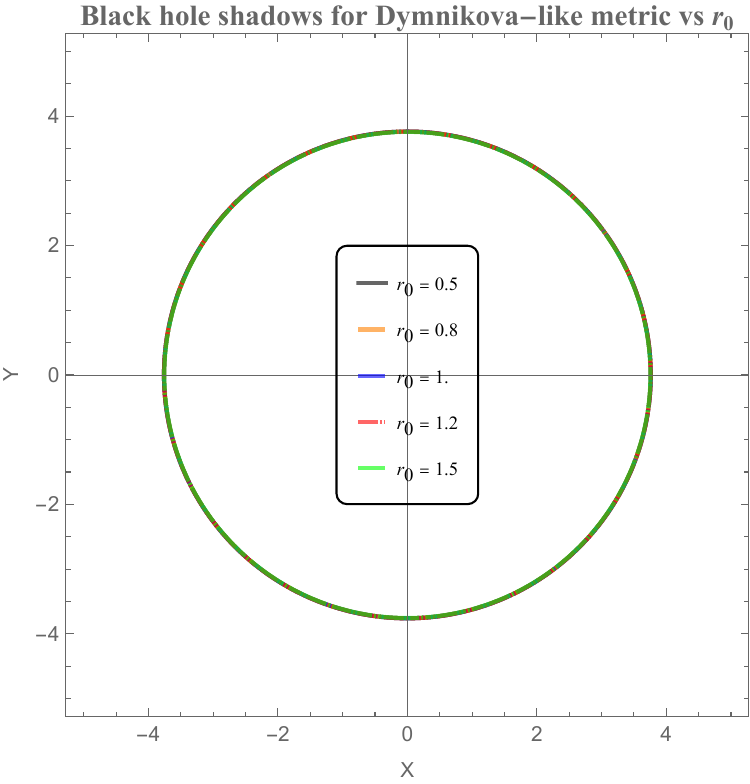}
\includegraphics[width=5.9cm,height=4.9cm]{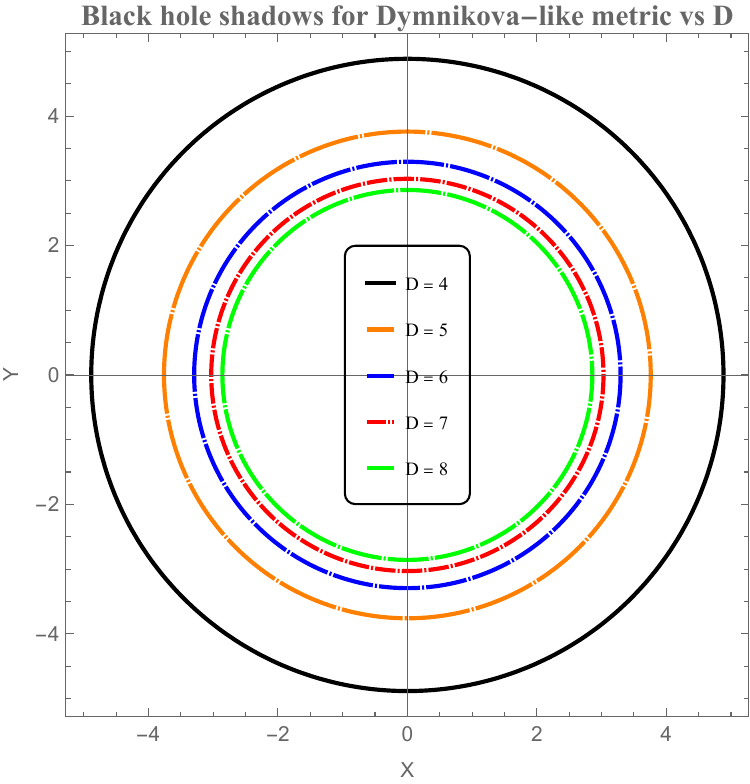}
\includegraphics[width=5.9cm,height=4.9cm]{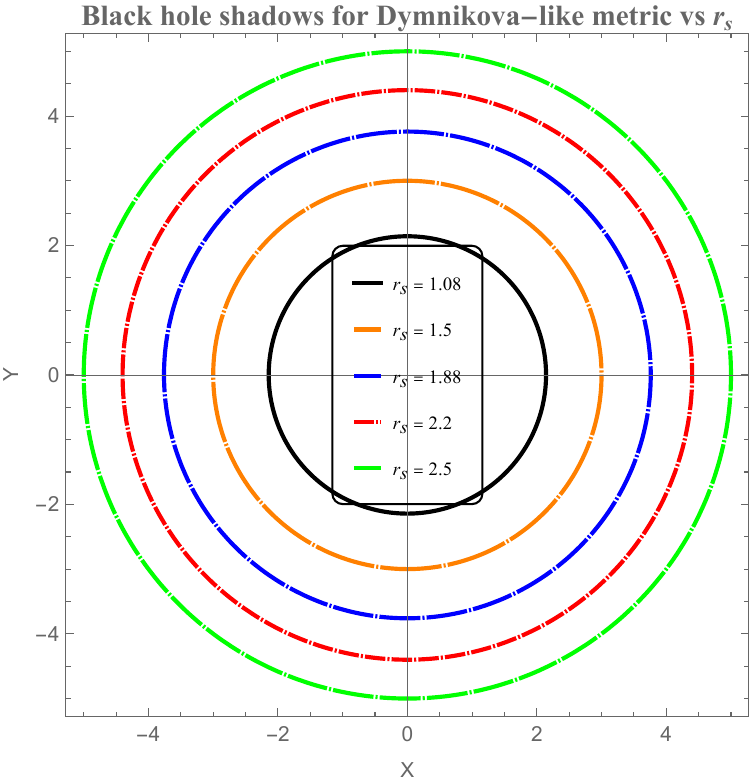}
\end{center}
\caption{\footnotesize The black hole shadow is studied under different parameter choices in higher-dimensional Dymnikova spacetimes. The left panel shows how varying $r_0$ modifies the profile while keeping $D=5$ and $r_s=1.88$ fixed. The middle panel illustrates the effect of changing the dimensionality $D$, with $r_0=1.0$ and $r_s=1.88$. The right panel depicts the influence of varying $r_s$ for fixed $r_0=1.0$ and $D=5$.}\label{Fig2}
\end{figure*}
%-----------------------------------------------------------------------------
%-----------------------------------------------------------------------------
\begin{figure*}
\begin{center}
\includegraphics[width=5.9cm,height=4.9cm]{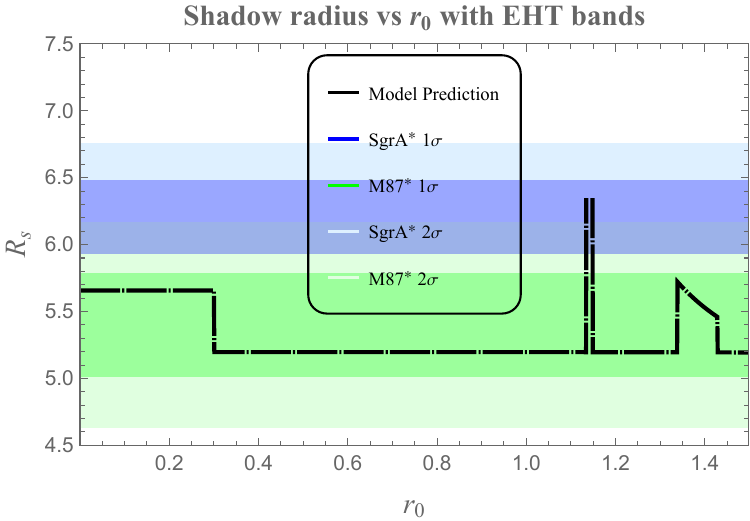}
\includegraphics[width=5.9cm,height=4.9cm]{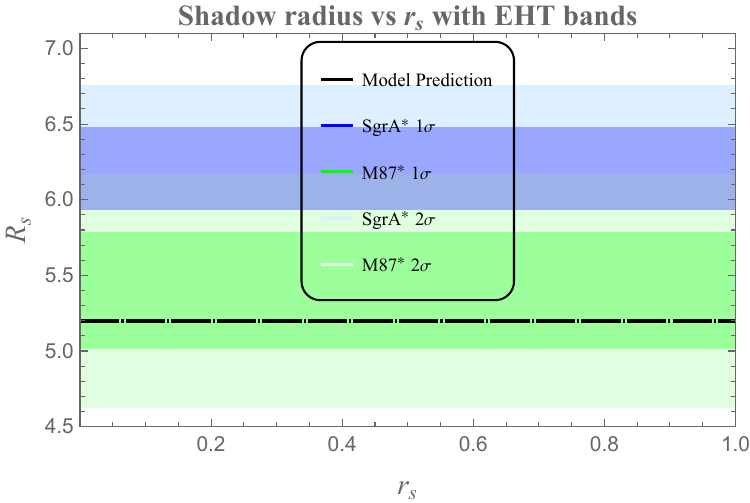}
\end{center}
\caption{\footnotesize The black hole shadow radius (solid black line) is illustrated in higher-dimensional Dymnikova spacetimes. The left panel shows the effect of varying $r_0$ while keeping $D=5$ and $r_s=1.88$ fixed, whereas the right panel demonstrates the impact of changing $r_s$ for fixed $r_0=1.0$ and $D=5$. The blue and green shaded regions correspond to the EHT horizon-scale observations of Sgr A$^*$ and M87$^*$ at the $1\sigma$ level, while the light blue and light green areas represent the $2\sigma$ consistency ranges, respectively. }\label{Fig3}
\end{figure*}
%-----------------------------------------------------------------------------
%-----------------------------------------------------------------------------
\begin{figure*}
\begin{center}
\includegraphics[width=5.9cm,height=4.9cm]{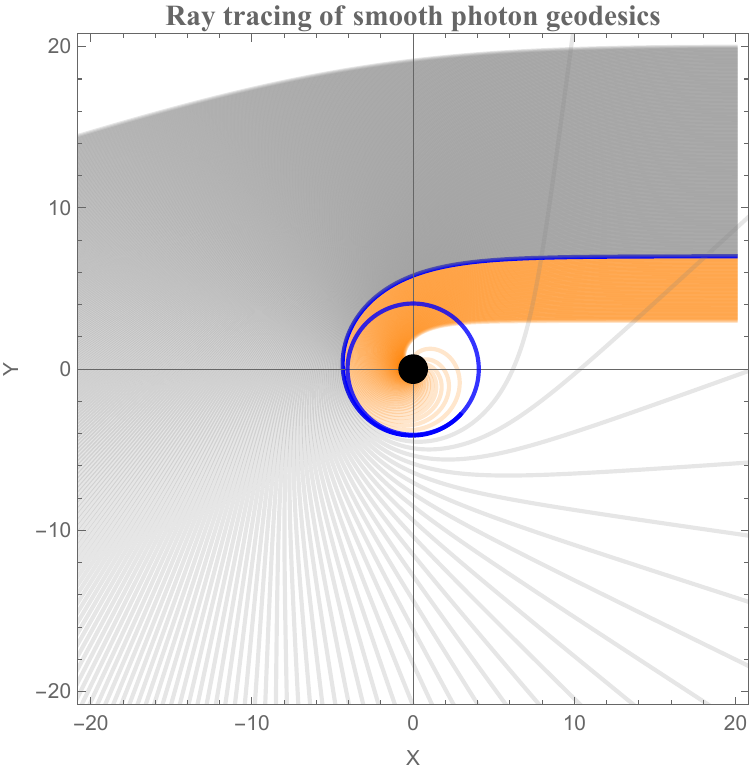}
\includegraphics[width=5.9cm,height=4.9cm]{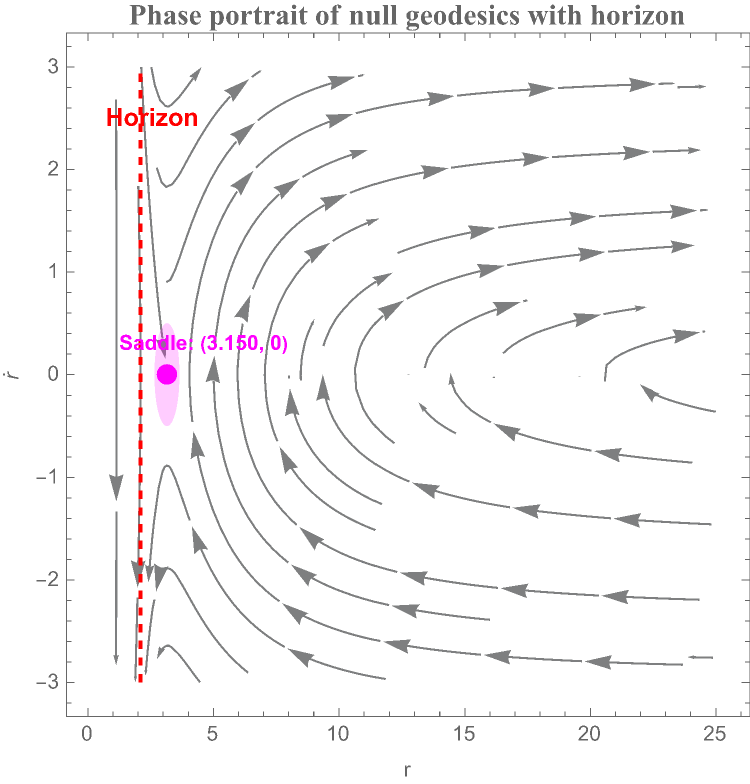}
\end{center}
\caption{\footnotesize The null geodesics and phase portrait in higher-dimensional Dymnikova spacetimes are analyzed. The left panel illustrates photon trajectories around the regular black hole, where the orange, blue, and gray curves correspond to different initial conditions, highlighting the distinction between capture, unstable circular orbits, scattering, respectively. The right panel shows the phase portrait of $\dot{r}$ versus $r$ for various choices of $r_0$ and $r_s$ with $D=5$. The magenta dot denotes the saddle (critical) point, which governs the stability properties of the photon sphere and separates the regions of bounded and unbounded motion.}\label{Fig4}
\end{figure*}
%---------------------------------------------------------------------------

To examine the behavior of the metric function in Eq. (\ref{laps}), we plot its variation with the radial coordinate $r$ in Fig. \ref{Fig1}. The graph reveals that, under the influence of the Dymnikova parameters and higher dimensions $D$, the spacetime can develop two distinct horizons: a Cauchy horizon $r_-$ and an event horizon $r_+$, particularly when $r_s \gg r_0$. Fig. \ref{Fig1} further illustrates how the lapse function $f(r)$ evolves for a spherically symmetric regular black hole as the parameters $r_0$, $D$, and $r_s$ vary. Specifically, as $r_0$ decreases or $r_s$ increases in $D=5$, and similarly when $D$ increases above four while keeping $r_0$ and $r_s$ fixed, the lapse function gradually drops, indicating a weakening of the gravitational potential near the black hole. This highlights the strong sensitivity of the spacetime geometry to changes in $r_0$, $r_s$, and $D$.

%--------------------------------------------
\textbf{Horizons and regular coordinates:}
%--------------------------------------------
The inner and outer horizons can be approximated by
\begin{equation}
r_{-} \approx r_0 \big[1 - O(e^{-r_0/r_s})\big], \quad r_{+} \approx r_s \big[1 - O(e^{-r_s^2/r_0^2})\big].
\end{equation}
Introducing Lema\^itre coordinates, the metric becomes regular at $r_\pm$ and near the origin:
\begin{equation}
ds^2 = -d\tau^2 + \frac{ r_s^{D-3} \big(1 - \mathcal{F}(r)\big)}{r^{D-3}} dr^2 + r^2 d\Omega_{D-2}^2,
\end{equation}
and in isotropic Eddington-Finkelstein coordinates:
\begin{equation}
ds^2 = -\bigg|1 - \frac{ r_s^{D-3} \big(1 - \mathcal{F}(r)\big)}{r^{D-3}}\bigg| dudv + r^2 d\Omega_{D-2}^2.
\end{equation}

%---------------------------
\textbf{Black hole mass:}
%---------------------------
The horizon condition $f(r_+) = 0$ gives
\begin{equation}
r_s^{D-3} = \frac{r_+^{D-3}}{1 - \exp\big(- r_+^{D-1}/r_{\star}^{D-1} \big)}.
\end{equation}
Using $r_{\star}^{D-1} = r_0^2 r_s^{D-3}$ with $r_0^2 = \frac{(D-1)(D-2)}{2 \rho_0}$, this can be written as a transcendental equation:
\begin{equation}
z e^z = - \frac{r_+^{D-1}}{r_0^2} , e^{-r_+^{D-1}/r_0^2}, \qquad z \equiv - \frac{r_+^{D-1}}{r_0^2 r_s^{D-3}}.
\end{equation}
Solving via the principal branch of the Lambert function $W_p$ yields
\begin{equation}
r_s^{D-3} = - \frac{r_+^{D-1}}{r_0^2} \frac{1}{ W_p\big(-\frac{r_+^{D-1}}{r_0^2} e^{-r_+^{D-1}/r_0^2}\big)}.
\end{equation}
The corresponding $D$-dimensional mass is
\begin{equation}
M = \frac{(D-2)\Omega_{D-2}}{16\pi} r_s^{D-3}, \qquad
\Omega_{D-2} = \frac{2 \pi^{(D-1)/2}}{\Gamma\big(\frac{D-1}{2}\big)},
\end{equation}
or explicitly in terms of $r_+$ and $r_0$:
\begin{eqnarray}
M = -\, \frac{(D-2)\,\Omega_{D-2}}{16 \pi} \; \frac{r_+^{\,D-1}}{r_0^2}\frac{1}{ W_p\Big(-\frac{r_+^{\,D-1}}{r_0^2} e^{-r_+^{\,D-1}/r_0^2}\Big)}.
\end{eqnarray}

\begin{table}[h!]
\centering
\caption{\footnotesize Hawking temperature $T_H$ for different parameter choices in higher-dimensional Dymnikova spacetimes.}
\scalebox{0.82}{
\begin{tabular}{c|c|c|c}
\hline
Parameter & Values & Fixed parameters & $T_H$ \\
\hline
$r_0$ & 0.8, 1.0, 1.2 & $D=5$, $r_s=1.88$ & 0.08571, 0.08960, 0.09476 \\
$D$   & 4, 5, 6       & $r_0=1.0$, $r_s=1.88$ & 0.04929, 0.08960, 0.13027 \\
$r_s$ & 1.5, 1.88, 2.0 & $r_0=1.0$, $D=5$ & 0.08569, 0.08960, 0.09124 \\
\hline
\end{tabular}}\label{tab0}
\end{table}

%-----------------------------------------------------------------------------
\textbf{Hawking radiation and energy emission in D Dimensions:}
%-----------------------------------------------------------------------------
Black holes continuously emit particles due to quantum effects near their horizons, leading to Hawking radiation and a gradual evaporation of the black hole. For a distant observer, this radiation appears to originate from a region roughly corresponding to the black hole's shadow, with an effective cross-section: $\sigma_{\rm lim} \simeq \pi R_s^2$ \cite{Wei:2013kza}, where $R_s$ depends on the photon's critical orbit and the observer's position \cite{Perlick:2021aok}. A fully consistent description of the energy emission must account for the D-dimensional phase space, spin of the particles, and the frequency-dependent greybody factors. The differential energy flux is therefore given by
\begin{eqnarray}
\frac{d^2 E}{d\omega dt} = \sum_{s} \sum_{\ell} N_s  \frac{\Gamma_\ell^s(\omega)}{e^{\omega/T_H(r_{+})} - (-1)^{2s}}  \frac{\omega^{D-2}}{(2\pi)^{D-1}}  \Omega_{D-2},
\end{eqnarray}
where $s$ labels the particle spin, $\ell$ denotes the angular mode, $N_s$ counts internal degrees of freedom, $\Gamma_\ell^s(\omega)$ are the greybody transmission coefficients, and $\Omega_{D-2}$ is the solid angle in $D-2$ dimensions. By introducing the absorption cross-section $\sigma_{\text{abs}}^s(\omega)$, which connects directly to the greybody (or transmission) coefficients, one can better understand how the emitted radiation deviates from the ideal blackbody spectrum due to spacetime curvature around the black hole. The relation is given by
\begin{eqnarray}
\sigma_{\text{abs}}^s(\omega) = \sum_{\ell} \frac{N_s \pi^{(D-2)/2}}{\Gamma\left(\frac{D-2}{2}\right)\omega^{D-2}} (2\ell + D - 3) \Gamma_\ell^s(\omega).
\end{eqnarray}
This expression accounts for the modification of the blackbody spectrum caused by gravitational effects near the event horizon. Accordingly, the energy emission spectrum can be written in an equivalent and compact form as
\begin{eqnarray}
\frac{d^2 E}{d\omega dt} = \sum_{s} N_s \frac{\sigma_{\text{abs}}^s(\omega) \Omega_{D-2}}{(2\pi)^{D-1}} \frac{\omega^{D-1}}{e^{\omega/T_H(r_{+})} - (-1)^{2s}}.
\end{eqnarray}

The Hawking temperature is determined by the surface gravity at the event horizon:
\begin{eqnarray}
T_H(r_{+}) = \frac{\kappa}{2\pi}, \qquad \kappa = \frac{1}{2} f'(r_+),
\end{eqnarray}
For the Dymnikova-type metric with lapse $f(r)$ defined in Eq.~(\ref{laps}), this temperature can be expressed explicitly as
\begin{eqnarray}
T_H(r_{+}) = \frac{1}{4\pi} \left[ \frac{D-3}{r_+} + \frac{D-1}{r_0^2 r_s^{D-3}} r_+ \, \exp\Big(- \frac{r_+^{D-1}}{r_0^2 r_s^{D-3}}\Big) \right].~~~~~
\end{eqnarray}

The impact of the parameters $r_0$, $D$, and $r_s$ on black hole thermodynamics is highlighted separately. Table~\ref{tab0} shows that the Hawking temperature $T_H(r_+)$ remains positive and rises as $r_0$, $D$, or $r_s$ increase, reflecting stronger thermal emission for larger black holes or in higher-dimensional spacetimes. Fig.~\ref{Fig6} illustrates the energy emission rate $\frac{d^2 E}{d\omega \, dt}$, which displays more intricate behavior. With $r_s$ and $D$ fixed, increasing $r_0$ from 0.01 to 1.0 (left panel) reduces the emission rate, suggesting that black holes with larger $r_0$ radiate less efficiently. On the other hand, holding $r_0$ and $r_s$ constant while increasing $D$ from 4 to 8 (middle panel), or fixing $D$ while increasing $r_s$ from 1.65 to 2.5 (right panel), boosts the emission rate, indicating more efficient radiation for higher dimensions or larger black holes. These findings underline that $r_0$, $D$, and $r_s$ play a key role in shaping both the Hawking temperature and the efficiency of energy emission, emphasizing how the geometry and dimensionality of higher-dimensional Dymnikova black holes govern their radiative behavior.

%----------------------------------------------------------
\textbf{Curvature invariants and singularity structure:}
%----------------------------------------------------------
To probe the nature of singularities, one can compute standard curvature invariants. The Ricci scalar reads
\begin{eqnarray}
R = g^{\mu\nu} R_{\mu\nu} &=& -\frac{1}{r^2} \Big( r^2 f''(r) + 2(D - 2) r f'(r)\nonumber\\ &&+ (D - 2)(D - 3) ( f(r) - 1 ) \Big),
\end{eqnarray}
the Ricci tensor squared is
\begin{eqnarray}
R_{\mu\nu} R^{\mu\nu} &=& +\frac{1}{2 r^4} \Big[ r^4 f''(r)^2 + 2(D - 2) r^2 f'(r)^2 + 2(D - 2)\nonumber\\ &&\times(D - 3) ( f(r) - 1 )^2 + 2(D - 2) r f'(r)\nonumber\\ && \times\big( r^2 f''(r) - 2 ( f(r) - 1 ) \big) \Big],
\end{eqnarray}
and the Kretschmann scalar reads
\begin{eqnarray}
K(r) = R^{\mu\nu\alpha\beta} R_{\mu\nu\alpha\beta} &=& +f''(r)^2 + \frac{2 (D - 2)}{r^2} f'(r)^2 \nonumber\\ && + \frac{2 (D - 2)(D - 3)}{r^4} ( f(r) - 1 )^2.~~~~~~~
\end{eqnarray}
At small $r$, the exponential term expands as
\begin{eqnarray}
e^{-r^{D-1}/r_\star^{,D-1}} \simeq 1 - \frac{r^{D-1}}{r_\star^{,D-1}} + \cdots,
\end{eqnarray}
so that
\begin{eqnarray}
f(r) \simeq 1 - \frac{r^2}{r_0^2} + \mathcal{O}(r^{D+1}),~
f'(r) \simeq -\frac{2r}{r_0^2},~
f''(r) \simeq -\frac{2}{r_0^2}.~~~~
\end{eqnarray}
Substituting these into the expressions above gives finite central values:
\begin{eqnarray}
R(r=0)&=&\frac{D(D-1)}{r_0^2}, \\
R_{\mu\nu}R^{\mu\nu}(r=0)&=&\frac{D(D-1)}{r_0^4}, \\
K(r=0)&=&\frac{2D(D-1)}{r_0^4}.
\end{eqnarray}
All curvature invariants remain finite at $r=0$; thus, the spacetime possesses a regular, de Sitter-like core and satisfies the defining property of a Dymnikova-type non-singular black hole.

%All these invariants diverge as $r \to 0$, confirming a true physical singularity at the black hole core.\\

%-----------------------------------------------------------------------------
\textbf{Geodesic trajectories around higher-dimensional Dymnikova black holes:}
%-----------------------------------------------------------------------------
To understand how black hole shadows emerge, one must first look at the paths followed by photons in curved spacetime. These null geodesics describe how light bends under gravity, and the collection of such trajectories outlines the dark silhouette, or shadow, that an observer would detect. In the case of higher-dimensional Dymnikova black holes, the geometry differs from the familiar Schwarzschild scenario. The Dymnikova profile regularizes the central singularity, and these modifications in the metric directly influence how photons move. As a result, the structure of the shadow carries signatures of the underlying geometry, making the study of geodesics the natural starting point. 

%---------------------------------------------------
\textbf{Lagrangian formulation of photon motion:}
%---------------------------------------------------
The motion of photons can be captured elegantly through a Lagrangian approach. For a general trajectory parameterized by an affine parameter $\tau$, the Lagrangian reads
\begin{equation}
\mathcal{L}(x,\dot{x}) = \tfrac{1}{2} g_{\mu\nu}(x)\dot{x}^\mu \dot{x}^\nu,
\end{equation}
where $\dot{x}^\mu = dx^\mu/d\tau$. Because photons travel along null geodesics, the condition $\mathcal{L} = 0$ applies, reflecting the fact that their worldlines have vanishing proper time. From this Lagrangian, the canonical momenta follow naturally:
%\begin{equation}
$p_\mu = \frac{\partial \mathcal{L}}{\partial \dot{x}^\mu}$.
%\end{equation}
These momenta are directly tied to conserved quantities whenever the geometry admits symmetries. For example, time-translation invariance ensures conservation of energy, while rotational symmetry guarantees conservation of angular momentum. Working in the standard coordinate system $x^\mu = (t, r, \theta, \phi)$, these constants of motion allow the geodesic equations to be reduced to an effective one-dimensional problem. The resulting radial equation highlights the conditions for circular photon orbits---the unstable trajectories that form the boundary of the shadow.

%--------------------------------------------------------------
\textbf{Equatorial photon motion and the black hole shadow:}
%--------------------------------------------------------------
To make the study of photon trajectories more manageable, we focus on motion restricted to the equatorial plane, $\theta = \pi/2$. This simplification is fully justified: the spherical symmetry of the black hole spacetime guarantees that any trajectory can be rotated into this plane without loss of generality. By doing so, the equations simplify considerably, while still capturing all the essential physics of photon motion and the resulting black hole shadow. The symmetries of the spacetime provide two fundamental constants of motion. Because the metric does not explicitly depend on the time coordinate $t$ or the azimuthal angle $\phi$, the corresponding Killing vectors lead to:
\begin{itemize}
    \item Photon energy $E$, associated with invariance under time translations.
    \item Angular momentum $L$, associated with invariance under rotations around the axis.
\end{itemize}

These conserved quantities make it possible to reduce the geodesic equations to a simpler form:
\begin{equation}
\dot{t} = \frac{E}{f(r)}, \qquad \dot{\phi} = \frac{L}{r^2},
\end{equation}
where $f(r)$ is the metric function of the higher-dimensional Dymnikova black hole. Here, the dot denotes differentiation with respect to the affine parameter $\tau$, which parametrizes the photon's path.

%---------------------------------------------
\textbf{Effective potential for photons:}
%---------------------------------------------
The radial motion can be recast in terms of an effective potential by imposing the null condition for massless particles:
%\begin{equation}
$g_{\mu\nu} \dot{x}^\mu \dot{x}^\nu = 0$.
%\end{equation}
This condition ensures that the photon's trajectory is lightlike. Using the conserved quantities, the radial equation becomes
\begin{equation}\label{reffpo}
\dot{r}^2 + V_{\rm eff}(r) = 0,~\text{with}~V_{\rm eff}(r) = f(r)\frac{L^2}{r^2} - E^2.
\end{equation}
%with
%\begin{equation}
%V_{\rm eff}(r) = f(r)\frac{L^2}{r^2} - E^2.
%\end{equation}

This formulation allows us to interpret photon motion as if it were a particle moving in a potential well: the shape of $V_{\rm eff}(r)$ determines whether photons fall into the black hole, escape to infinity, or remain trapped in circular orbits.

%---------------------------------------------------
\textbf{Radial trajectories and photon orbits:}
%---------------------------------------------------
Expressed in terms of the azimuthal angle $\phi$, the radial evolution along a photon trajectory reads
\begin{equation}
\frac{dr}{d\phi} = \pm r \sqrt{f(r) \left(\frac{r^2 E^2}{L^2 f(r)} - 1\right)}.
\end{equation}
This equation describes how the photon's radial position changes as it orbits the black hole. The impact parameter $b = L/E$ naturally arises from this expression and plays a key role in determining the photon's fate: whether it falls into the black hole, escapes, or hovers near the photon sphere.

%----------------------------------------------
\textbf{Photon sphere and the shadow edge:}
%----------------------------------------------
The black hole shadow is defined by photons that asymptotically approach unstable circular orbits at the photon sphere radius $r_p$. Mathematically, these orbits satisfy
\begin{equation}\label{Vrp}
V_{\rm eff}(r_p) = 0, \qquad \left. \frac{dV_{\rm eff}}{dr} \right|_{r=r_p} = 0,
\end{equation}
or, equivalently, $\left. \frac{dr}{d\phi} \right|_{r=r_p} = 0$. Plugging these conditions back into the orbital equation gives
\begin{equation}
\frac{dr}{d\phi} = \pm r \sqrt{f(r) \left(\frac{r^2 f(r_p)}{r_p^2 f(r)} - 1\right)}.
\end{equation}

Only photons with precisely the right impact parameter can orbit at $r_p$. Slightly smaller impact parameters lead to capture by the black hole, while slightly larger ones escape to infinity. This delicate balance produces the sharply defined boundary of the shadow observed from afar.

%-----------------------------------------------------------------
\textbf{Angular size and apparent shape of the black hole shadow:}
%-----------------------------------------------------------------
For a static observer located at a radius $r_0$, the angle $\beta$ between the observer's line of sight and the radial direction determines how the black hole appears on the celestial sphere. The trajectory of photons reaching the observer satisfies
\begin{equation}
\cot\beta = \left. \frac{\sqrt{g_{rr}}}{\sqrt{g_{\phi\phi}}} \frac{dr}{d\phi} \right|_{r=r_0}.
\end{equation}
This relation can be rearranged to express the angular size of the shadow in terms of the metric and the radius $r_p$ of the circular photon orbit:
%\begin{equation}
$\sin^2 \beta = \frac{f(r_0) r_p^2}{r_0^2 f(r_p)}$.
%\end{equation}
In the limit where the observer is very far from the black hole ($r_0 \to \infty$), the angular radius simplifies to
\begin{equation}
\sin\beta \simeq \frac{r_p}{\sqrt{f(r_p)}}.
\end{equation}
This shows that the observed shadow is directly determined by the photon sphere, the unstable circular orbit of photons around the black hole.

%---------------------------------------------------------
\textbf{Celestial coordinates and shadow projection:}
%---------------------------------------------------------
To describe the shadow's apparent shape on the observer's image plane, one introduces celestial coordinates $X$ and $Y$, defined in a plane perpendicular to the line of sight at $\theta_0 = \pi/2$. In the asymptotic limit $r_0 \to \infty$, these coordinates are expressed as
\begin{align}
X &= \lim_{r_0 \to \infty} \left. -r_0^2 \sin\theta_0 \frac{d\phi}{dr} \right|_{(r_0, \theta_0)}, \\
Y &= \lim_{r_0 \to \infty} \left. r_0^2 \frac{d\theta}{dr} \right|_{(r_0, \theta_0)}.
\end{align}
These coordinates provide a two-dimensional projection of the shadow, encoding both its size and shape as seen by a distant observer. The photon sphere radius $r_p$ is determined by the effective potential $V_{\rm eff}(r)$ for photon motion, satisfying the conditions in (\ref{Vrp}). Once $r_p$ is known, the corresponding shadow radius is
\begin{equation}\label{eqRs}
R_s = \frac{r_p}{\sqrt{f(r_p)}}.
\end{equation}
This relation highlights that the shadow is a direct imprint of the spacetime geometry, with the metric function $f(r)$ encoding the gravitational influence of the black hole.

%-------------------------------------------------------------------------------------

%----------------------------------------------
\textbf{Conditions for the photon sphere:}
%----------------------------------------------
The photon sphere exists at a radius $r_p$ that satisfies
\begin{equation}
r \frac{df(r)}{dr} - 2 f(r) = 0.
\end{equation}
Substituting the explicit form of the metric function $f(r)$ in the presence of matter distributions and corrections gives the equation
\begin{widetext}
\begin{eqnarray}
(D-1)\frac{16 \pi M}{(D-2) \Omega_{D-2} r^{D-3}}\left[1 - \exp\Big(-\frac{r^{D-1} (D-2) \Omega_{D-2}}{16 \pi M r_0^2}\Big)\right] - \frac{D-1}{r_0^2} r^2 \exp\Big(-\frac{r^{D-1} (D-2) \Omega_{D-2}}{16 \pi M r_0^2}\Big) -2= 0,~~~
\end{eqnarray}
\end{widetext}
which can be solved to find $r_p$ and thus the shadow radius $R_s$. This shows that the black hole shadow is sensitive not only to the mass $M$ but also to the structure of the higher-dimensional Dymnikova spacetime and any additional modifications encoded in the metric. Since the photon sphere radius cannot be expressed analytically, it must be computed numerically, after which the shadow radius is obtained directly from relation (\ref{eqRs}). Fig. \ref{Fig2} presents how the shadow radius evolves when the Dymnikova profile parameters and the dimensionality of spacetime are varied. The analysis reveals systematic trends: when the scale parameter $r_0$ varies between $0.5$ and $1.5$, the shadow radius remains essentially unchanged for the case $D=5$. In contrast, increasing the number of dimensions from $D=4$ to $D=8$ leads to a steady decrease in the shadow size, while variations in the Schwarzschild radius $r_s$ within the range $1.08$ to $2.5$ for $D=5$ result in a gradual growth of the shadow radius. These behaviors arise from the delicate interplay between the Dymnikova regularization parameters and the dimensional structure of spacetime. The parameter $r_0$ controls the degree of smoothing of the central core, while $r_s$ encodes the effective gravitational radius. When combined with higher-dimensional corrections, these quantities reshape the effective gravitational potential that photons experience. A stronger effective pull increases the tendency of photons to be captured in unstable circular orbits, enlarging the shadow. Conversely, a weaker curvature reduces light bending, allowing photons to escape more easily, thereby shrinking the shadow. From an observational perspective, this means that the shadow size is not fixed solely by mass, but rather by the fine-tuning between $r_0$, $r_s$, and $D$. In higher dimensions, the gravitational field is effectively diluted, softening the photon capture region and reducing the black hole's apparent imprint on the observer's cosmic backdrop. The joint effect of the Dymnikova profile and the dimensional extension therefore acts to regulate the strength of spacetime curvature, directly shaping the geometry and visibility of the shadow.

%-----------------------------------------------------------------------------
\begin{table}[h!]
\centering
\caption{\footnotesize {Allowed ranges of the Dymnikova black hole parameters $r_0$ and $r_s$, where $r_0$ sets the characteristic scale of the nonsingular de Sitter-like core and $r_s$ denotes the effective Schwarzschild radius associated with the black hole mass, consistent with the EHT shadow radius measurements of SgrA* and M87* at the $1\sigma$ and $2\sigma$ confidence levels.}}
\scalebox{0.82}{
\begin{tabular}{c|c|c|c|c}
\hline
Parameter & SgrA* 1$\sigma$ & SgrA* 2$\sigma$ & M87* 1$\sigma$ & M87* 2$\sigma$ \\
\hline
$r_0^{\rm range}$ & 0.0062 -- 0.119 & 0.00001 -- 0.208 & 0.035 -- 0.084 & 0.00001 -- 0.125 \\
$r_s^{\rm range}$ & 0.0048 -- 0.102 & 0.00001 -- 0.182 & 0.029 -- 0.077 & 0.00001 -- 0.112 \\
\hline
\end{tabular}}\label{tab2}
\end{table}
%-----------------------------------------------------------------------------

%-----------------------------------------------------------------------------
\begin{table}[h!]
\centering
\caption{\footnotesize {Shadow radius $R_s$ as a function of the Dymnikova black hole parameters $r_0$ and $r_s$, where $r_0$ controls the scale of the nonsingular de Sitter-like core and $r_s$ denotes the effective Schwarzschild radius associated with the black hole mass, shown together with the EHT observational constraints. Legend: $\checkmark$ = within the observationally allowed range, $\times$ = outside the observational range.}}
\scalebox{0.85}{
\begin{tabular}{c|c|c|c|c|c|c}
\hline
Parameter & Value & $R_s$ & Sgr A* 1$\sigma$ & M87* 1$\sigma$ & Sgr A* 2$\sigma$ & M87* 2$\sigma$ \\
\hline
\multicolumn{7}{c}{Varying $r_0$ (0.00001 $\rightarrow$ 0.208)} \\
\hline
$r_0$ & 0.00001 & 6.50 & $\times$ & $\times$ & $\checkmark$ & $\checkmark$ \\
$r_0$ & 0.050   & 6.20 & $\checkmark$ & $\times$ & $\checkmark$ & $\checkmark$ \\
$r_0$ & 0.100   & 5.90 & $\checkmark$ & $\checkmark$ & $\checkmark$ & $\checkmark$ \\
$r_0$ & 0.150   & 5.55 & $\times$ & $\checkmark$ & $\checkmark$ & $\checkmark$ \\
$r_0$ & 0.200   & 5.30 & $\times$ & $\checkmark$ & $\checkmark$ & $\checkmark$ \\
\hline
\multicolumn{7}{c}{Varying $r_s$ (0.00001 $\rightarrow$ 0.182)} \\
\hline
$r_s$ & 0.00001 & 6.45 & $\times$ & $\times$ & $\checkmark$ & $\checkmark$ \\
$r_s$ & 0.040   & 6.10 & $\checkmark$ & $\times$ & $\checkmark$ & $\checkmark$ \\
$r_s$ & 0.080   & 5.80 & $\checkmark$ & $\checkmark$ & $\checkmark$ & $\checkmark$ \\
$r_s$ & 0.120   & 5.50 & $\times$ & $\checkmark$ & $\checkmark$ & $\checkmark$ \\
$r_s$ & 0.160   & 5.25 & $\times$ & $\checkmark$ & $\checkmark$ & $\checkmark$ \\
\hline
\end{tabular}}\label{tab3}
\end{table}

%---------------------------------------
\textbf{Observational Implications:}
%--------------------------------------
To explore constraints on black hole parameters, we turn to the landmark observations provided by the EHT collaboration \cite{EventHorizonTelescope:2019dse, EventHorizonTelescope:2019ggy}. These measurements offer an exceptional opportunity to confront theoretical models with actual astrophysical data and to establish meaningful bounds on the parameters that characterize the black hole environment. For this analysis, we focus on two well-studied supermassive black holes: Sgr A$^{*}$ at the center of the Milky Way and M87$^{*}$ in the Virgo cluster. Both sources are well-suited for this study because they can be approximated as static, spherically symmetric systems, which aligns with the assumptions of our theoretical framework. By comparing the predicted angular size of the black hole shadow with the angular diameters measured by the EHT, we can derive upper limits on the parameters $r_0$ and $r_s$. This comparison relies on three observational inputs: the shadow's angular diameter $\theta$, the distance to the source $D$, and the black hole mass $M$. The latest EHT data provide precise values for these quantities for both Sgr A$^{*}$ and M87$^{*}$, enabling a direct translation from observations to constraints on our model parameters.

Recent results from the EHT have provided accurate determinations of the shadow sizes, distances, and masses for the two best-studied supermassive black holes, M87$^{*}$ and Sgr A$^{*}$ \cite{EventHorizonTelescope:2019dse, EventHorizonTelescope:2019ggy}. The reported observational values are:
$ \theta_{M87^{*}} = 42 \pm 3 ~\mu\text{as},~ 
\theta_{Sgr A^{*}} = 48.7 \pm 7 ~\mu\text{as},~
D_{M87^{*}} = 16.8 \pm 0.8 ~\text{Mpc},~
D_{Sgr A^{*}} = 8277 \pm 9 \pm 33 ~\text{pc},~
M_{M87^{*}} = (6.5 \pm 0.7)~ 10^{9} M_{\odot},~
M_{Sgr A^{*}} = (4.297 \pm 0.013)~10^{6} M_{\odot}$. To connect these measurements with theory, one introduces the dimensionless shadow diameter per unit mass, $d_{sh} = \frac{D \, \theta}{M}$. For the two black holes, this relation yields
\begin{eqnarray}
d^{M87^{*}}_{sh} = (11 \pm 1.5)M,
\quad 
d^{Sgr A^{*}}_{sh} = (9.5 \pm 1.4)M.
\end{eqnarray}

What these results reveal is that the apparent shadow scales smoothly with the black hole's mass and distance, in close agreement with theoretical expectations. In practice, this means the EHT observations provide a direct and reliable way to turn astrophysical data into quantitative constraints on black hole models. Observational data provide key constraints on the parameters $r_0$ and $r_s$ associated with higher-dimensional Dymnikova spacetimes. These limits are shown in Fig.~\ref{Fig3}, with $r_0$ on the left and $r_s$ on the right. By incorporating the latest EHT observations of Sgr A$^{*}$ and M87$^{*}$, we can establish precise bounds on these parameters. For Sgr A$^{*}$, the supermassive black hole at the center of our galaxy, $r_0$ lies between 0.0062 and 0.119 at the $1\sigma$ confidence level, widening to 0.00001-0.208 at $2\sigma$. The corresponding limits for $r_s$ are 0.0048-0.102 ($1\sigma$) and 0.00001-0.182 ($2\sigma$). In contrast, M87$^{*}$, which exhibits a larger shadow, allows slightly higher values: $r_0$ ranges from 0.035 to 0.084 ($1\sigma$) and 0.00001-0.125 ($2\sigma$), while $r_s$ spans 0.029-0.077 ($1\sigma$) and 0.00001-0.112 ($2\sigma$).

These results show that M87$^{*}$ consistently permits larger values of both $r_0$ and $r_s$ compared with Sgr A$^{*}$, reflecting the effect of higher-dimensional Dymnikova structures on spacetime geometry. Accordingly, the comparatively larger shadow of M87$^{*}$ aligns naturally with these parameter ranges. Table~\ref{tab2} provides a clear summary of the derived bounds, while Table~\ref{tab3} lists the allowed intervals alongside the corresponding shadow radii for both black holes under different parameter configurations. In summary, the observations confirm that M87$^{*}$ accommodates somewhat larger parameter values, providing new insight into the influence of higher-dimensional Dymnikova structures on spacetime geometry.

%------------------------------------------------------------------------
\textbf{Photon trajectories and impact of higher-dimensional parameters:} The propagation of photons around the black hole depends sensitively on the parameters $r_{0}$, $D$, and $r_{s}$. By solving the orbital equation numerically, we obtain the photon trajectories shown in Fig.~\ref{Fig4} (left panel), for $r_{0}=-0.01$, $D=5$, and $r_{s}=1.647$, with all other parameters fixed. The shape of the orbits is largely determined by the impact parameter, $\tilde{b}=L/E$, which measures the photon's angular momentum per unit energy. Small values of $\tilde{b}$ cause photons to fall directly into the horizon, whereas larger values lead to grazing paths that experience only mild bending. All orbits swirl anti-clockwise, consistent with the chosen orientation of angular momentum. Several features are apparent in Fig.~\ref{Fig4}. The impact parameter acts as an effective radial coordinate for distant observers, controlling the bending angle. As $\tilde{b}$ increases, deflection grows, diverging at the critical value $\tilde{b}=\tilde{b}_{ps}$, which defines the photon sphere. Photons at this radius orbit the black hole along unstable circular paths, with small perturbations sending them either into the horizon or back to infinity (shown by blue curves). For even larger $\tilde{b}$, the bending angle decreases, producing the sequence of trajectories highlighted by the color transition orange $\rightarrow$ blue $\rightarrow$ gray. The photon sphere sets the edge of the black hole shadow and determines the scale of strong lensing. In Schwarzschild spacetime, it occurs at $r_{p}=3M$ with $\tilde{b}_{ps}=3\sqrt{3}M$. In the present case, Fig.~\ref{Fig4} shows that the additional structure introduced by $r_{0}$, $D$, and $r_{s}$ modifies both the photon sphere radius and the critical impact parameter, with orbits gradually approaching the Schwarzschild limit as $\tilde{b}$ grows. Physically, $r_0$, $D$, and $r_{s}$ regulate the balance between local matter distribution and higher-dimensional effects, shaping photon motion through spacetime curvature. These trajectories illustrate not only the geometry of null geodesics but also the observable impact of higher-dimensional Dymnikova parameters on black hole shadows and lensing patterns, providing a clear connection to light propagation in strong gravity regions.

\textbf{Assessing null geodesic stability through a phase space approach:} The stability of null circular geodesics is analyzed in the $(r,\dot{r})$ phase space, where equilibrium occurs at $(r_c,0)$ with $\dot{r}=0$; differentiating Eq.~(\ref{reffpo}) yields $\ddot{r} = -\frac{dV_{\rm eff}(r)}{dr}$, and with variables $a_1=\dot{r}$ and $a_2=\ddot{r}$, the system becomes
\begin{equation}
a_1=\dot{r}, \quad a_2=-\frac{dV_{\rm eff}(r)}{dr},
\end{equation}
whose Jacobian is
$\mathcal{J}=\begin{pmatrix}0 & 1 \\ -V_{\rm eff}''(r) & 0\end{pmatrix}$, leading to the secular equation $|\mathcal{J}-\lambda I|=0$ with eigenvalue squared: $\lambda_L^2=-V_{\rm eff}''(r)$, so that the Lyapunov exponent $\lambda_L$ measures the divergence or convergence of trajectories, identifying unstable saddles when $\lambda_L^2>0$ and stable centers when $\lambda_L^2<0$ \cite{goldhirsch1987stability, Yang:2023hci}. Fig.~\ref{Fig4} (right panel) displays the phase flow of circular null geodesics in  higher-dimensional Dymnikova spacetimes for different $(r_0, r_s)$, showing that photon sphere orbits are dynamically unstable: even infinitesimal perturbations drive photons either into the horizon or to infinity. The saddle point at $(3.150, 0)$ marks this instability, with smaller $r_s$ (for $D=5$, $r_0=0.01$, $L=12$) shifting the critical radius inward, steepening the potential and amplifying the sensitivity of photon trajectories to disturbances.

\textbf{Concluding Remarks:} In this work, we systematically investigated higher-dimensional Dymnikova black holes in detail, highlighting both their physical properties and observational aspects within regular black hole scenarios. Starting from a density profile inspired by the gravitational analogue of the Schwinger mechanism, we constructed the corresponding metric and extended it to arbitrary dimensions $D$, maintaining spherical symmetry in the stress-energy tensor. The lapse function was obtained explicitly, showing how the parameters $r_0$, $r_s$, and $D$ shape the spacetime geometry. We found that the metric can feature two distinct horizons---a Cauchy horizon $r_-$ and an event horizon $r_+$---especially when $r_s \gg r_0$. Regular coordinate systems, such as Lema\^{i}tre and isotropic Eddington-Finkelstein coordinates, were introduced to remove coordinate singularities at $r_\pm$ and near the origin. The black hole mass in $D$ dimensions was expressed in terms of the outer horizon $r_+$ and the characteristic scale $r_0$, using the Lambert function to solve the horizon condition.

Building on this geometric construction, we explored thermodynamic properties through the Hawking temperature $T_H$ and the corresponding energy emission rate. Our findings show that $T_H$ remains positive across the parameter space and generally rises with $r_0$, $r_s$, or $D$, indicating stronger thermal emission for larger or higher-dimensional black holes. The energy emission rate behaves more subtly: increasing $r_0$ lowers the emission rate, while higher $D$ or larger $r_s$ increases it, highlighting the sensitivity of black hole radiation to geometry and dimensionality. The regularity of the geometry was confirmed through curvature invariants. The Ricci scalar, Ricci tensor squared, and Kretschmann scalar remain finite for $r>0$, verifying the absence of physical singularities inside the Dymnikova core. This supports the interpretation of the solution as a regular black hole, smoothly connecting a de Sitter-like core to asymptotic Schwarzschild behavior.

Extending our analysis to the motion of light, photon motion was examined using a Lagrangian approach for null geodesics. Focusing on equatorial motion, we identified conserved energy and angular momentum, reducing the dynamics to an effective one-dimensional problem. The effective potential governs radial motion, circular photon orbits, and the black hole shadow boundary. The photon sphere radius $r_p$ and the corresponding shadow radius $R_s$ were computed numerically for different values of $r_0$, $r_s$, and $D$. We found that variations in $r_0$ have minimal effect on the shadow size for $D=5$, while increasing $D$ reduces $R_s$, and larger $r_s$ leads to a gradual increase in the shadow. This illustrates how the Dymnikova regularization, the Schwarzschild radius, and dimensionality together shape photon capture and the observable shadow. Finally, by projecting photon trajectories onto celestial coordinates, we determined the apparent shape and angular size of the shadow for a distant observer. The shadow radius depends directly on the photon sphere and the metric function, reflecting both the Dymnikova profile and higher-dimensional effects. Systematic changes in $r_0$, $r_s$, and $D$ show that the shadow size is influenced not only by mass but also by regularization and dimensional parameters, underlining how geometry and dimensionality together determine observational signatures.

{ When we compare our results with previous studies on regular black holes and higher-dimensional models, it becomes clear that our work provides complementary insights into the Dymnikova profile and the influence of spacetime dimensionality. Studies of higher-dimensional black holes---including Schwarzschild-Tangherlini, Einstein-Maxwell-Kaluza-Klein, Gauss-Bonnet, Myers-Perry rotating solutions, and regular higher-dimensional spacetimes---show that both the number of dimensions and the scale of the central core or higher-curvature corrections leave clear marks on horizon structure, thermodynamics, photon spheres, and shadow sizes. For example, Tangherlini \cite{Tangherlini:1963bw} and Kastor et al. \cite{Kastor:2010gq} explored how static higher-dimensional Schwarzschild-like solutions shape thermodynamic properties and horizon behavior, while Boulware \& Deser \cite{Boulware:1985wk}, Cai \cite{Cai:2001dz}, and Zhang et al. \cite{Zeng:2020dco} demonstrated that Gauss-Bonnet couplings can alter photon spheres, shadows, and phase structures in higher dimensions. Myers \& Perry \cite{Myers:1986un, Myers:1986rx} and Emparan \& Reall \cite{Emparan:2001wn} showed that rotation in higher dimensions produces new horizon geometries, black rings, and distinctive optical features that depend on spacetime dimensionality. Meanwhile, Lobos et al. \cite{Lobos:2024fzj} and Bueno et al. \cite{Bueno:2024dgm} studied regular higher-dimensional solutions, highlighting that larger regularization scales tend to reduce shadow sizes and modify emission rates, effects that could be observed near the horizon. In this context, our results indicate that the Dymnikova regularization parameter $r_0$ and the spacetime dimension $D$ should fall within ranges compatible with Schwarzschild-Tangherlini-like shadow sizes and expected thermodynamic behavior. Interestingly, while increasing $D$ generally reduces the shadow radius, consistent with earlier studies, the relatively weak dependence on $r_0$ for $D=5$ emphasizes a distinctive feature of the Dymnikova model, producing more subtle observational effects compared with other regular black holes. Altogether, these findings underscore that both the number of dimensions and the specifics of regularization or higher-order curvature corrections are crucial for predicting observable signatures---such as photon capture, shadow structure, and thermodynamic behavior---in higher-dimensional black hole spacetimes.
}

Overall, this study offers a detailed characterization of higher-dimensional Dymnikova black holes, connecting their internal structure, horizon properties, thermodynamics, and photon dynamics to the shadows they produce. It establishes a clear link between the theoretical construction of regular black holes and their potential astrophysical signatures in extended spacetimes, providing a framework for future exploration of higher-dimensional regular black hole solutions.

\textbf{Acknowledgements:} This research was funded by the Science Committee of the Ministry of Science and Higher Education of the Republic of Kazakhstan (Grant No. AP22682760).  

\textbf{Conflict Of Interest statement: } The authors declare that they have no known competing financial interests or personal relationships that could have appeared to influence the work reported in this paper.

\textbf{Data Availability Statement:}  This manuscript has no associated data, or the data will not be deposited. (There is no observational data related to this article. The necessary calculations and graphic discussion can be made available on request.)

\bibliography{references}

\end{document}